\documentclass[10pt]{article}
\usepackage{latexsym}
\usepackage{amssymb}
\usepackage{amsfonts}
\usepackage{amsmath}
\usepackage{theorem}

\newtheorem{theorem}{Theorem}
\newtheorem{lemma}{Lemma}
\newtheorem{corollary}{Corollary}
\newtheorem{definition}{Definition}

\newtheorem{proposition}{Proposition}

\begin{document}

\title{Self-Averaging Identities \\
for Random Spin Systems}
\author{Luca De Sanctis
\footnote{ICTP, Strada Costiera 11, 34014 Trieste, Italy,
{\tt<lde\_sanc@ictp.it>}} ,
Silvio Franz
\footnote{ICTP, Strada Costiera 11, 34014 Trieste, Italy,
{\tt<franz@ictp.it>}}}

\maketitle

\begin{abstract}
We provide a systematic treatment of self-averaging identities
for various spin systems. The method is quite
general, basically not relying on the nature of the model,
and as a special case recovers the Ghirlanda-Guerra 
and Aizenman-Contucci identities,
which are therefore proven, together with their extension, 
to be valid in a vaste class of spin
models. We use the dilute spin glass as a guiding
example.
\end{abstract}

\noindent{\em Key words and phrases:} spin glasses, diluted 
spin glasses, Ghirlanda-Guerra, self-averaging.


\section{Introduction}

Despite many years of intense work, and the much awaited proof of the
validity of the Parisi ansatz for the free-energy of the 
Sherrington-Kirkpatrick (SK) and
related models, the mathematical comprehension of thermodynamics of
mean field spin glasses remains largely incomplete. We know from
theoretical physics that in fully connected models, all the properites
of the low temperature spin glass phase can be encoded in the
probability distribution of the overlap between two different copies of
the system. The analysis of Parisi et al. predicts an ultrametric
organization of the phases (see \cite{mpv} and references therein).
So far the rigorous proof (or disproof) of ultrametricity, and, more in
general, the analysis of the structure of Gibbs measures at low
temperature, turned out to be a very difficult task. A step in this
direction was performed by Ghirlanda and Guerra in \cite{gg}. They found
a simple and elegant way, based on the self-averaging of the
internal energy, to prove a remarkable property of the overlaps. Given
$s$ replicas, the Gibbs measure must be such that
when one adds a further replica this is either
identical to one these, or statistically independent of them; each
case occurring with the same probability. More
generally, various constraints on the distribution of the different
overlaps have been found in the same spirit (\cite{ac, parisi}). Such
features have found several applications (\cite{talabook, bovierbook})
in the rigorous analysis of spin glass models. For example, the
property of non-negativity of the overlap, which in some models plays
a role in turning the cavity free-energy into a rigorous lower bound,
turns out to be a consequence of the Ghirlanda-Guerra self-averaging
identities (\cite{talabook}). In the same way these identity have
a role in the rigorous analysis of spin glasses close to the critical
temperature (\cite{abds}).

In more general spin-glass systems, like finite dimensional systems or
spin systems on random graphs, the statistics of the overlap are not
enough to fully characterize the low temperature spin glass phase. For
instance, in diluted models the statistics of the local cavity
fields, or equivalently of all the multi-overlaps, is necessary to
describe the low temperature thermodynamic properties. In this
paper, we analyse two families of identities for the local fields and
multi-overlap distributions that are a consequence of self-averaging
relations. We will see that one of the two families is
a consequence of the self-averaging with respect to the 
Gibbs measure or, equivalently, 
of stochastic stability, as the two phenomena turn out to be 
equivalent. The other family of identities is instead a consequence 
of self-averaging with respect to the global measure
(quenched after Gibbs). Our conclusions will not rely much 
on the specific form of the Hamiltonian of the model. 
We will however use the example of spin
models on sparse random graphs (dilute spin glass models), where we
expect that our results could provide hints for progresses in the mathematical
analysis of the low temperature phases. Diluted mean field spin
glasses have, in recent time, attracted a lot of attention in
statistical physics, due to their intrinsic interest of spin glasses
where each spin interacts with a finite number of variables, but more
importantly because fondamental problems in computer science, such as
the random K-SAT and graph coloring, the random X-OR-SAT, tree
reconstruction \cite{mm1} and others, admit a formulation in
terms of spin glass systems on random graphs. 
The cavity approach to these problems has led in
many cases to results believed to be exact, albeit for the moment
several rigorous proofs are still lacking. 

Some of the identities that we will discuss appeared already in
\cite{flt} to discuss free energy bounds in diluted models with
non-Poissonian connectivity. Here we re-derive with different methods
this family of identities, and we exhibit a second family of new identities.


\section{The notations}

We will use the stereotypical dilute spin glass model,
the Viana-Bray (VB), to introduce here the notations we need,
and to derive our results in the next two sections.

Notations:
$\alpha, \beta$ are 
non-negative real numbers
(degree of connectivity and inverse temperature
respectively);
$P_\zeta$ is a Poisson random variable of mean $\zeta$;
$\{i_\nu\}, \{j_\nu\}$, etc. are independent identically 
distributed random variables, 
uniformly distributed over the points $\{1,\ldots, N\}$;
$\{J_\nu\}, J$, etc. are
independent identically distributed 
random variables, with symmetric distribution;
$\mathcal{J}$ is the set of all the quenched 
random variables  above; the map
$\sigma: i \rightarrow \sigma_{i},\ i\in\{1,\ldots , N\}$ 
is a spin configuration from the 
configuration space $\Sigma=\{-1,1\}^{N}$;
$\pi_{\zeta}(\cdot)$ is the Poisson measure
of mean $\zeta$; 
$\mathbb{E}$ is an average over all (or some of) the 
quenched variables;
$\omega_\mathcal{J}$ or simply $\omega$ is the Bolztmann-Gibbs 
average explicitly written below;
$\Omega_N$ or simply $\Omega$ are a product of the needed 
number of independent identical copies (replicas) 
of $\omega_\mathcal{J}$;
$\langle\cdot\rangle$ will indicate the composition 
of an $\mathbb{E}$-type average over 
quenched variables and the
Boltzmann-Gibbs average over the spin variables (see below).
We will often drop the dependance on 
some variables or indices or slightly
change notations to lighten the expressions, 
when there is no ambiguity.
As a main example, consider
the Hamiltonian of the Viana-Bray model, defined as
\begin{equation*}
\label{ham}
H^{VB}_N(\sigma, \alpha; \mathcal{J})=
-\sum_{\nu=1}^{P_{\alpha N}} J_\nu \sigma_{i_\nu}\sigma_{j_\nu}\ .
\end{equation*}
We will limit to the case $J=\pm 1$, without loss of 
generality \cite{gt1}.
We follow the usual basic definitions and notations 
of thermodynamics for the partition function $Z_{N}$,
the pressure $p_{N}$,
the free energy per site $f_{N}$ and its thermodynamic limit $f$,
so to have in general
\begin{equation*}
Z_{N}(\beta,\alpha)=Z(H_{N}; \beta,\alpha)=\sum_{\{\sigma\}}
\exp(-\beta H_N(\sigma,\alpha))\ ,
\end{equation*}
\begin{equation*}
p_{N}(\beta,\alpha)=-\beta f_N(\beta,\alpha)=\frac1N \mathbb{E}
\ln Z_N(\beta,\alpha)\ ,\ f(\beta,\alpha)=\lim_{N\to\infty} f_N(\beta,\alpha)\ .
\end{equation*}
The Boltzmann-Gibbs average of an observable 
$\mathcal{O}:\Sigma\to\mathbb{R}$ is
\begin{equation*}
\omega(\mathcal{O})=
Z_N(\beta,\alpha)^{-1}
\sum_{\{\sigma\}}\mathcal{O}(\sigma)\exp(-\beta
H_N(\sigma,\alpha))\ ,
\end{equation*}
$\mathbb{E}$ denotes the average with respect to the quenched variables,
and $\langle\cdot\rangle=\mathbb{E}\omega(\cdot)$ is the
global average.

The multi-overlaps $q_{1\cdots m}:\Sigma^{m}\to[-1,1]$,
where we use the notation 
$\Sigma^{n}=\Sigma^{(1)}\times\cdots\times\Sigma^{(n)}$, 
among the ``replicas'' 
$\Sigma^{(r_{1})}\ni\sigma^{(r_{1})},\ldots,\Sigma^{(r_{n})}\ni\sigma^{(r_{n})}$ 
is defined by
\begin{equation*}
\label{overlap}
q_{r_1\cdots r_n}=\frac{1}{N}\sum_{i=1}^N\sigma_i^{(r_1)}
\cdots\sigma_i^{(r_n)}\ ,
\end{equation*}
but sometimes we will just write $q_n$; $q_{1}$ can be identified with
the magnetization $m$
$$
m=\frac{1}{N}\sum_{i=1}^N\sigma_i\ .
$$
Dealing with binary spins, we will not be using powers of the spins,
so we will often drop the brackets $()$ in the replica index for 
the spins, so that $\sigma^{s}_{i}\equiv\sigma^{(s)}_{i}$ will mean the $i$-th spin 
from the replica $s$, $\Sigma^{(s)}$, not the $s$-th power of 
$\sigma_{i}$.
Notice
\begin{equation}
\label{q}
\mathbb{E}\omega^{2n}
(\sigma_{i_{.}})=\langle q_{1\cdots 2n}\rangle\ ,\ 
\mathbb{E}\omega
(\sigma_{i_{.}})=\mathbb{E}\omega
(m)=\langle m \rangle\ .
\end{equation}


\section{Stochastic Stability and self-averaging of the Gibbs measure}
\label{qs}

In the study of finite connectivity models it emerged that
in a suitable propability space it is possible to
formulate an exact variational principle for the computation
of the free energy. This was obtained with the introduction
of Random Multi-Overlap Structures (RaMOSt). We refer
to \cite{lds1} for details.
The ROSt approach is based on the use of generic random weights
to average the ``cavity'' part and the relative ``internal
correction'' in the free energy (these are the numerator and
the denominator of the trial free energy $G_N$ introduced in
(\ref{gtfdilute}). See \cite{lds1} for details).
Here we are not interested in a detailed discussion of
the RaMOSt approach, but we study the effect of a 
perturbation to the measure of our model, which does not
need to be the Gibbs measure. That is why introduce this
more general weighting scheme, although the reader may
keep in mind the Gibbs measure as a guiding example.

\subsection{Random Multi-Overlap Structures}

The proper framework for the calculation of the free energy per spin
is that of the Random Multi-Overlap Structures (RaMOSt, see
\cite{lds1} for more details).
\begin{definition}
  Given a probability space $\{\Omega,\mu(d\omega)\}$,
  a {\bf Random Multi-Overlap Structure}
  $\mathcal{R}$ is a triple
  $(\tilde{\Sigma}, \{\tilde{q}_{2n}\}, \xi)$ where
  \begin{itemize}
  \item $\tilde{\Sigma}$ is a discrete space;
  \item $\xi: \tilde{\Sigma}\rightarrow\mathbb{R}_+$
    is a system of random weights, such that $\sum_{\gamma\in\tilde{\Sigma}}
    \xi_\gamma\leq\infty$ $\mu$-almost surely;
  \item $\tilde{q}_{2n}:\tilde{\Sigma}^{2n}\rightarrow\mathbb{R},
    n\in\mathbb{N}$ is a positive semi-definite \emph{Multi-Overlap
      Kernel} (equal to 1 on the diagonal of $\tilde{\Sigma}^{2n}$, so
    that by Schwartz inequality $|\tilde{q}|\leq 1$).
  \end{itemize}
\end{definition}
A RaMOSt needs to be equipped with $N$ independent copies
of a random field
$\{\tilde{h}^{i}_{\gamma}(\alpha; \tilde{J})\}_{i=1}^{N}$ and
with another random field
$\hat{H}_{\gamma}(\alpha; \hat{J})$
such that
\begin{eqnarray}
\frac{d}{d\alpha}\mathbb{E}\ln \sum_{\gamma\in\tilde{\Sigma}}\xi_{\gamma}
\exp(-\beta\tilde{h}^{i}_{\gamma})
& = & 2\sum_{n>0}\frac{1}{2n}\tanh^{2n}(\beta)
(1-\langle \tilde{q}_{2n}\rangle)\ , \label{eta} \\
\frac{d}{d\alpha}\mathbb{E}\ln \sum_{\gamma\in\tilde{\Sigma}}\xi_{\gamma}
\exp(-\beta \hat{H}_{\gamma})
&=& \sum_{n>0}\frac{1}{2n}\tanh^{2n}(\beta)
(1-\langle\tilde{q}^{2}_{2n}\rangle)\ .\label{kappa}
\end{eqnarray}
These two fields are employed in the definition of the
trial pressure
\begin{equation}\label{gtfdilute}
G_{N}(\mathcal{R};\beta)=\frac 1N\mathbb{E}\ln
\frac{\sum_{\gamma, \sigma}\xi_{\gamma}
\exp(-\beta\sum_{i=1}^{N}\tilde{h}^{i}_{\gamma}
\sigma_{i})}{\sum_{\gamma}\xi_{\gamma}
\exp(-\beta \hat{H}_{\gamma})}\ .
\end{equation}
The reason why this is the proper framework for the 
calculation of the free energy is explained 
by the next \cite{lds1}
\begin{theorem}[Extended Variational Principle]
Taking the infimum for each $N$
separately of the trial function $G_N(\mathcal{R};\beta)$ over the space
of all RaMOSt's, the resulting sequence tends to the limiting pressure
$-\beta f(\beta)$ of the VB model as $N$ tends to infinity:
\begin{equation*}
\label{evp-d}
-\beta f(\beta)=\lim_{N\rightarrow\infty}
\inf_{\mathcal{R}}G_{N}(\mathcal{R};\beta)\ .
\end{equation*}
\end{theorem}
A RaMOSt $\mathcal{R}$ is said to be optimal if
$G(\mathcal{R};\beta)=-\beta f(\beta)\ \ \forall\ \beta$.
We will denote by $\Omega$ the measure
associated to the RaMOSt weights $\xi$ as well.

The Boltzmann RaMOSt \cite{lds1} is optimal, and
constructed by thinking of a reservoir of $M$ spins $\tau$
$$
\Sigma=\{-1,1\}^M\ni\tau\ ,\ \xi_\tau=\exp(-\beta H_M(\tau))\ ,\
\tilde{q}_{1\cdots 2n}=\frac1M\sum_{k=1}^M\tau^{(1)}_k\cdots\tau^{(2n)}_k
$$
with
$$
\tilde{h}^{i}_{\tau}(\alpha)=\sum_{\nu=1}^{P_{2\alpha}}
\tilde{J}_{\nu}^{i}\tau_{k_{\nu}^{i}}\ , \
\hat{H}_{\tau}(\alpha N)=
-\sum_{\nu=1}^{P_{\alpha N}}
\hat{J}_\nu \tau_{k_\nu}\tau_{l_\nu}
$$
and $\tilde{J},\hat{J}$ all independent copies of $J$.

Let $c_{i}=2\cosh(\beta\tilde{h}^{i})$.
It is possible to show \cite{lds1} that
optimal RaMOSt's enjoy the same
factorization property enjoyed by the Boltzmann
RaMOSt and described in the next \cite{lds1}
\begin{theorem}[Factorization of optimal RaMOSt's]
\label{lisboa-d}
With the possible exception of a zero measure set
of values of the degree of connectivity, the following Ces{\`a}ro  limit
is linear in $N$ and $\bar{\alpha}$
\begin{equation*}\label{limrost}
\mathbf{C}\lim_{M}\mathbb{E}\ln\Omega_M
\{c_{1}\cdots c_{N}\exp[-\beta \hat{H}(\bar{\alpha})]\}
=N(-\beta f +\alpha A)+\bar{\alpha}A\ ,
\end{equation*}
where
\begin{equation}\label{defa}
A=\sum_{n=1}^{\infty}\frac{1}{2n}
\mathbb{E}\tanh^{2n}(\beta J)(1-\langle q_{2n}^{2}\rangle)\ .
\end{equation}
\end{theorem}
This factorization property is called {\sl invariance with 
respect to the cavity step},
or {\sl Quasi-Stationarity},
and it is found in the hierarchical Parisi ansatz as well.
When $\bar{\alpha}$ is zero, the theorem above states
the factorization of the cavity fields, and it is possible to show that
from this property one can deduce the family of identities we will
discuss in the next subsection \cite{bds2}. When one removes instead
the cavity terms $c_{1},\ldots , c_{N}$ from the previous theorem,
the statement becomes what is usually referred to as Stochastic
Stability. We will show that the latter too implies the same family
of identities. We will have in mind the case of a small perturbation
of our spin system, but what we find holds for more
general RaMOSt's, provided the previous theorem holds, that is
for Quasi-Stationary RaMOSt's.

\subsection{The first family of identities}

We will now prove a lemma that expresses the stability
of the Gibbs measure of our model against a macroscopic but
small stochastic perturbation. In different terms, the lemma expresses
the linear response of the free energy to the connectivity shift
the perturbation consists of.
The lemma we are about to prove will be used to show that
from stochastic stability one can deduce a certain self-averaging
which in turn imposes a family of constraints on the distribution of the
overlaps.
\begin{lemma}\label{lemma}
  Let $\Omega\ ,\ \langle\cdot\rangle$ be the usual Gibbs and quenched
  Gibbs expectations at inverse temperature $\beta$, associated with
  the Hamiltonian $H_{N}(\sigma, \alpha; \mathcal{J})$. Then, with the
  possible exception of a zero measure set of values of the degree
  of connectivity,
\begin{equation}\label{stability}
\lim_{N\to\infty}\mathbb{E}\ln\Omega\exp\bigg(\beta^{\prime}
\sum_{\nu=1}^{P_{\alpha^{\prime}}}
J^{\prime}_{\nu}\sigma_{i^{\prime}_{\nu}}\sigma_{j^{\prime}_{\nu}}\bigg)
=\alpha^{\prime}\sum_{n=1}^{\infty}\frac{1}{2n}
\tanh^{2n}(\beta^{\prime})(1-\langle q^{2}_{2n}\rangle)\ ,
\end{equation}
where the random 
variables $P_{\alpha^{\prime}}, \{J^{\prime}_{\nu}\}$,
$\{i^{\prime}_{\nu}\},\{j^{\prime}_{\nu}\}$ are independent
copies of the analogous random variables 
in the Hamiltonian in contained in $\Omega$.
\end{lemma}
Notice that, in distribution 
\begin{equation}
\label{continuo1}
\beta\sum_{\nu=1}^{P_{\alpha N}}
J_{\nu}\sigma_{i_{\nu}}\sigma_{j_{\nu}}+\beta^{\prime}
\sum_{\nu=1}^{P_{\alpha^{\prime}}}
J^{\prime}_{\nu}\sigma_{i^{\prime}_{\nu}}\sigma_{j^{\prime}_{\nu}}
\sim\beta\sum_{\nu=1}^{P_{(\alpha+\alpha^{\prime}/N)N}}
J^{\prime\prime}_{\nu}\sigma_{i_{\nu}}\sigma_{j_{\nu}}
\end{equation}
where $\{J^{\prime\prime}_{\nu}\}$ are independent copies 
of $J$ with probability $\alpha N/(\alpha N+\alpha^{\prime})$ and 
independent copies of $J\beta^{\prime}/\beta$ with probability
$\alpha^{\prime}/(\alpha N+\alpha^{\prime})$. In the right hand
side above, the quenched random variables will be collectively denoted
by $\mathcal{J}^{\prime\prime}$.
Notice also that 
the sum of Poisson random variables 
is a Poisson random variable with
mean equal to the sum of the means, and hence we can write
\begin{equation}
  \label{continuo2}
  A_{t}\equiv\mathbb{E}\ln\Omega\exp\bigg(\beta^{\prime}
  \sum_{\nu=1}^{P_{\alpha^{\prime}t}}
  J^{\prime}_{\nu}\sigma_{i^{\prime}_{\nu}}\sigma_{j^{\prime}_{\nu}}\bigg)
  =\mathbb{E}\ln\frac{Z_{N}(\alpha_{t};
    \mathcal{J}^{\prime\prime})}{Z_{N}(\alpha;\mathcal{J})}\ ,
\end{equation}
where we defined, for $t\in[0, 1]$,
\begin{equation}
  \label{continuo3}
  \alpha_{t}=\alpha+\alpha^{\prime}\frac{t}{N}
\end{equation}
so that $\alpha_{t}\rightarrow\alpha\ \forall \ t$ as $N\to\infty$.\\
\textbf{Proof}.
Let us compute the $t$-derivative of $A_{t}$, as defined in 
(\ref{continuo2})
\begin{equation*}
  \frac{d}{dt}A_{t}=
  \mathbb{E}\sum_{m=1}^{\infty}\frac{d}{dt}\pi_{\alpha^{\prime} t}(m)
  \ln\sum_{\sigma}\exp\bigg(\beta^{\prime}\sum_{\nu=1}^{m}
  J^{\prime}_{\nu}\sigma_{i^{\prime}_{\nu}}
  \sigma_{j^{\prime}_{\nu}}\bigg)\ .
\end{equation*}
Using the following
elementary property of the Poisson measure
\begin{equation}\label{poisson}
  \frac{d}{dt}\pi_{t\zeta}(m)=\zeta(\pi_{t\zeta}(m-1)-\pi_{t\zeta}(m))
\end{equation}
we get
\begin{eqnarray*}
  \frac{d}{dt}A_{t}&=&
  \alpha^{\prime}\mathbb{E}\sum_{m=0}^{\infty}
  [\pi_{\alpha^{\prime} t}(m-1)
  -\pi_{\alpha^{\prime} t}(m)]
  \ln\sum_{\sigma}\exp(\beta^{\prime}\sum_{\nu=1}^{m}J^{\prime}_{\nu}
  \sigma_{i^{\prime}_{\nu}}\sigma_{j^{\prime}_{\nu}})\\
  {}&=&\alpha^{\prime}\mathbb{E}\ln\sum_{\sigma}
  \exp(\beta^{\prime} J^{\prime}\sigma_{i^{\prime}_{m}}
  \sigma_{j^{\prime}_{m}})
  \exp(\beta^{\prime}\sum_{\nu=1}^{P_{\alpha^{\prime} t}}J^{\prime}_{\nu}
  \sigma_{i^{\prime}_{\nu}}\sigma_{j^{\prime}_{\nu}})\\
  {}&{}&\hspace{3cm}-\alpha^{\prime}\mathbb{E}\ln\sum_{\sigma}
  \exp(\beta^{\prime}\sum_{\nu=1}^{P_{\alpha^{\prime} t}}J^{\prime}_{\nu}
  \sigma_{i^{\prime}_{\nu}}\sigma_{j^{\prime}_{\nu}})\\
  {}&=&\alpha^{\prime}\mathbb{E}\ln\Omega_{t}
  \exp(\beta^{\prime} J^{\prime}\sigma_{i^{\prime}_{m}}
  \sigma_{j^{\prime}_{m}})\ ,
\end{eqnarray*}
where we included the $t$-dependent weights
in the average $\Omega_{t}$.
Now use the following identity
$$
\exp(\beta^{\prime} J^{\prime}\sigma_{i}\sigma_{j})
=\cosh(\beta^{\prime} J^{\prime})+\sigma_{i}\sigma_{j}
\sinh(\beta^{\prime} J^{\prime})
$$
to get
\begin{equation*}
  \frac{d}{dt}A_{t}
  =\alpha^{\prime}\mathbb{E}\ln\Omega_{t}
  [\cosh(\beta^{\prime} J^{\prime})(1+\tanh(\beta^{\prime} 
  J^{\prime})\sigma_{i^{\prime}_{m}}
  \sigma_{j^{\prime}_{m}})]\ .
\end{equation*}
It is clear that
\begin{equation*}
  \mathbb{E}\ \omega_{t}^{2n}(\sigma_{i_{m}}\sigma_{j_{m}})
  =\langle q^{2}_{2n}\rangle_{t}\ ,
\end{equation*}
so we now expand the logarithm in power series and see that,
in the limit of large $N$, as $\alpha_{t}\to\alpha$
the result does not depend on $t$,
everywhere the expectation $\langle\cdot\rangle_{t}$
is continuous
as a function of the parameter $t$
(or equivalently as a function of the degree of
connectivity). 
From the comments that preceded
the current proof, formalized 
in (\ref{continuo1})-(\ref{continuo2})-(\ref{continuo3}), 
this is the same as assuming that
$\Omega$ is regular as a function of $\alpha$,
because
$J^{\prime\prime}\to J$ in the sense that in the large
$N$ limit $J^{\prime\prime}$ can only take
the usual values $\pm 1$ since the probability of being
$\pm \beta^{\prime}/\beta$ becomes zero. 
Therefore integrating
over $t$ from 0 to 1 is the same as multiplying by 1.
Due to the symmetric distribution of $J$,
the expansion of the logarithm yields the right hand side of
(\ref{stability}),
where the odd powers are missing. $\Box$

Let us define
$$
\hat{H}(\alpha^{\prime};\mathcal{J})=\sum_{\nu=1}^{P_{\alpha^{\prime}}}
J_{\nu}\sigma_{i_{\nu}}\sigma_{j_{\nu}}\sim H(\alpha^{\prime}/N;\mathcal{J})
$$
Let us now consider the statement of Lemma \ref{lemma},
in the case of two independent perturbations (the quenched variables
in the perturbations, denoted by $\mathcal{J}^{\prime}_{1},
\mathcal{J}^{\prime}_{2}$, are independent one another and independent 
from those in the Hamiltonian
of the Boltzmann factor).
Then
the fundamental theorem of calculus can be used twice
to extend the statement of the previous lemma to
\begin{equation}\label{twopert}
\mathbb{E}\ln\Omega[\exp(-\beta^{\prime}_1\hat{H}
(\alpha^{\prime}_{1};\mathcal{J}_{1}^{\prime})
-\beta^{\prime}_2\hat{H}(\alpha^{\prime}_{2};
\mathcal{J}_{2}^{\prime}))]=(\alpha^{\prime}_{1}
+\alpha^{\prime}_{2})A\ ,
\end{equation}
where $A$ again does not depend,
in the thermodynamic limit, on 
$\alpha_1^{\prime},\alpha_2^{\prime}$, and incidentally has 
the same form as the right hand side of
(\ref{stability}). In the equation above,
assumed to be taken in the thermodynamic limit, $\Omega$ 
is the Gibbs measure associated with the unperturbed
Hamiltonian of the original model, and the same holds
for the averages appearing in $A$, just like in the previous lemma.
Clearly we then have (omitting the dependence on the 
independent quenched random variables)
$$
\frac{\partial^{2}}{\partial\alpha^{\prime}_{1}
  \partial\alpha^{\prime}_{2}}
\mathbb{E}\ln\Omega[\exp(-\beta^{\prime}_{1}\hat{H}(\alpha^{\prime}_{1})
-\beta^{\prime}_{2}\hat{H}(\alpha^{\prime}_{2}))]=0\ ,
$$
and again in the thermodynamic limit $\Omega$
does not include any perturbation with 
$\alpha_1^{\prime},\alpha_2^{\prime},\beta_1^{\prime},\beta_2^{\prime}$.
A simple computation yields
\begin{multline*}
  \frac{\partial^{2}}{\partial\alpha^{\prime}_{1}
    \partial\alpha^{\prime}_{2}}
  \mathbb{E}\ln\Omega[\exp(-\beta^{\prime}_1\hat{H}(\alpha^{\prime}_{1})
  -\beta^{\prime}_2\hat{H}(\alpha^{\prime}_{2}))]=0\\
  =\mathbb{E}\ln\Omega[\exp(\beta^{\prime}_{1}J^{\prime}_{1}
  \sigma_{i_{1}}\sigma_{j_{1}}+
  \beta^{\prime}_{2}J^{\prime}_{2}\sigma_{i_{2}}\sigma_{j_{2}}]\\
  -\mathbb{E}\ln\Omega[\exp(\beta^{\prime}_{1}J^{\prime}_{1}
  \sigma_{i_{1}}\sigma_{j_{1}}]\Omega[\exp(\beta^{\prime}_{2}
  J^{\prime}_{2}
  \sigma_{i_{2}}\sigma_{j_{2}}]
\end{multline*}
Every time a derivative with respect to a pertubing parameter is taken, 
the relative perturbation is added to the weights of the measure
$\Omega$, but if the pertubation is small (like in our case, as
explained in the previous lemma) it disappears from the measure
in the thermodynamic limit. This is true for almost
all values of the perturbing parameters.
Hence we may assume that both in the equation above and
in the next calculation $\beta_1^{\prime},\beta_2^{\prime}$ are
not in the measure $\Omega$, and we get
\begin{multline}\label{recap}
  \frac{\partial^{2}}{\partial(\beta^{\prime}_{1}J_{1})
    \partial(\beta^{\prime}_{2}J_{2})}
  \mathbb{E}\ln\Omega[\exp(\beta^{\prime}_{1}J^{\prime}_{1}
  \sigma_{i_{1}}\sigma_{j_{1}}+
  \beta^{\prime}_{2}J^{\prime}_{2}\sigma_{i_{2}}\sigma_{j_{2}}]\\
  =\mathbb{E}\Omega(\sigma_{i_{1}}\sigma_{j_{1}})
  -\mathbb{E}\Omega(\sigma_{i_{1}})\Omega(\sigma_{j_{1}})=0\ ,
\end{multline}
at the price of a zero measure set of values of the parameters
(which allows us to use always the unperturbed expectation 
$\Omega$).
The first line of this equation gives us the 
generator of a family of relations that
we will obtain by means of an expansion in powers
of $\beta_1^{\prime},\beta_2^{\prime}$.
The second line of the equation formulates the self-averaging
(with respect to the Gibbs measure)
implied by the stochastic stability.

So we proceed starting from the next lemma and the next theorem,
summarizing what we just discussed. 
\begin{lemma}\label{prima} Let $\Omega^{\prime}$
be the Gibbs measure including two independent perturbations
of the form 
$$
\hat{H}(\alpha^{\prime})=\sum_{\nu=1}^{P_{\alpha^{\prime}}}
J^{\prime}_{\nu}\sigma_{i_{\nu}}\sigma_{j_{\nu}}
$$
with parameters $\alpha_1^{\prime},\alpha_2^{\prime}, 
\beta_1^{\prime},\beta_2^{\prime}$ like in (\ref{twopert}).
Then, recalling that $m$ is the magnetization, 
the following self-averaging (with respect to the Gibbs measure)
identity
\begin{equation}\label{phi}
  \lim_{N\to\infty}\mathbb{E}\{\Omega^{\prime}
  (m^{2})
  -[\Omega^{\prime}(m)]^{2}\}=0 
\end{equation}
holds for almost all values of the two perturbing parameters
$\alpha_1^{\prime},\alpha_2^{\prime}$.
\end{lemma}
We will see again that in the first line of equation (\ref{recap})
the expression remains zero even without the derivative.
In fact the generator of the identities we want to prove
is expressed in the following 
\begin{theorem}\label{thm-generatore} 
In the thermodynamic limit the following holds
for almost all values of $\alpha_1^{\prime}$ and 
$\alpha_2^{\prime}$:
\begin{eqnarray}\label{eq-generatore}
&& \mathbb{E}\ln 
\Omega^{\prime}(\exp(\beta_1^{\prime} 
J^{\prime}_1\sigma_{i_1}\sigma_{j_1}
+\beta_2^{\prime} J^{\prime}_2\sigma_{i_2}\sigma_{j_2}))= \\
\nonumber
&&\mathbb{E}\ln 
\Omega^{\prime}(\exp(\beta_1^{\prime}
J_1^{\prime}\sigma_{i_1}\sigma_{j_1}))
+\mathbb{E}\ln 
\Omega^{\prime}(\exp(\beta_2^{\prime}
J_2^{\prime}\sigma_{i_2}\sigma_{j_2}))\ .
\end{eqnarray}
\end{theorem}
The relations we will derive are a simple consequence
of this theorem, and fomalized in the next
\begin{corollary} In the thermodynamic limit, 
for almost all values of the perturbing parameters 
$\alpha_1^{\prime},\alpha_2^{\prime}$ we have
$$
\sum_{a=0}^{\min\{r,s\}}(-)^{a+1}\frac{(2r+2s-a-1)!}{a!(2r-a)!(2s-a)!}
\langle q^2_{2r}q^2_{2s}\rangle^{\prime}_a=0  
\ \ \forall\ r,s \in \mathbb{N}\ ,
$$
where the subscript a in the global average
$\langle\cdot\rangle^{\prime}_a=\mathbb{E}\Omega^{\prime}_{a}$ 
means that $a$ replicas are in common
among those in $q_{r}$ and those in $q_{s}$,
so that in particular $\Omega_{a}$ is (in a given term)
the product measure of 
only $2r+2s-a$ copies of $\omega^{\prime}$.
\end{corollary}
The ``prime''
superscript indicates as usual that the measure contains the perturbations,
which vanish in the thermodynamic limit but allows
us ``almost sure'' statements only.

{\bf Proof}. The following shorthand will be employed
$$
t_1=\tanh(\beta^{\prime}_1 J^{\prime}_1)\ ,\ 
t_2=\tanh(\beta^{\prime}_2 J^{\prime}_2)\ ,
$$
$$
\ \Omega_1=\Omega^{\prime}(\sigma_{i_1}
\sigma_{j_1})\ ,\
\Omega_2=\Omega^{\prime}(\sigma_{i_2}\sigma_{j_2})\ , \ 
\Omega_{12}=\Omega^{\prime}(\sigma_{i_1}
\sigma_{j_1}\sigma_{i_2}\sigma_{j_2})
$$ 
and
$$
W=\Omega^{\prime}(\exp
(\beta_1^{\prime} J^{\prime}_1\sigma_{i_1}\sigma_{j_1}
+\beta_2^{\prime} J^{\prime}_2\sigma_{i_2}\sigma_{j_2}))\ ,
$$
Observe that, if we let $\delta=1,2$,
\begin{equation}
\label{derivative}
\frac{\partial}{\partial {\beta J^{\prime}_\delta}}
=(1-t_{\delta}^2)\frac{\partial}{\partial {t_{\delta}}}\ .
\end{equation}
Now,
$$
\ln W =
\ln(1+t_1\Omega_1+t_2\Omega_{2}+t_1t_2\Omega_{12})+
\ln\cosh\beta J^{\prime}_1+\ln\cosh\beta J^{\prime}_2
$$
and
\begin{multline*}
\ln(1+t_1\Omega_1+t_2\Omega_{2}+t_1t_2\Omega_{12})=\\
\sum_{n=1}^{\infty}\sum_{l=0}^{n}\sum_{m=0}^{l}\frac{(-)^{n+1}}{n}
\binom{n}{l}\binom{l}{m}
t_1^{n-l+m}t_2^{n-m}\Omega_{1}^{m}\Omega_2^{l-m}\Omega_{12}^{n-l}\\
=\sum_{n,l,m}(-)^{n+1}\frac{(n-1)!}{(n-l)!(l-m)!m!}
t_1^{n-l+m}t_2^{n-m}\Omega_{1}^{m}\Omega_2^{l-m}\Omega_{12}^{n-l}\ .
\end{multline*}
The derivatives in (\ref{recap}) kill the two terms with the
hyperbolic cosines, and from (\ref{derivative}) we know that we can
replace the derivatives with respect to $\beta J^{\prime}_\delta$ with
the derivatives with respect to $t_\delta$, $\delta=1,2$.  Notice that
the logarithm just expanded is zero for $t_{1}=0$ and for $t_{2}=0$,
therefore as its derivative like in (\ref{recap}) is zero, the
logarithm itself is zero. This is why Theorem \ref{thm-generatore}
holds, being (\ref{eq-generatore}) just the integral of the second
line in (\ref{recap}).

Thanks to (\ref{q}), if we put
$$
n-l+m=r\ ,\ n-m = s\ ,\ n-l = a
$$
we get
$$
\sum_{r,s}\mathbb{E}[t_1^{r}t_2^{s}]\sum_{a=0}^{\min\{r,s\}}(-)^{a+1}
\frac{(r+s-a-1)!}{a!(r-a)!(s-a)!}
\langle q^{2}_{r}q^{2}_{s}\rangle^{\prime}_a=0 
$$
where $\langle\cdot\rangle_a$ means that $a$ replicas are in common
among those in $q_{r}$ and those in $q_{s}$. 
Hence the statement of the theorem to be proven
$$
\sum_{a=0}^{\min\{2r,2s\}}(-)^{a+1}\frac{(2r+2s-a-1)!}{a!(2r-a)!(2s-a)!}
\langle q^{2}_{2r}q^{2}_{2s}\rangle^{\prime}_a=0\ .
$$

\subsection{Generalization to smooth functions of multi-overlaps}

The fact  that in our formulas we  always got the square  power of the
overlaps  is  due  to  the   fact  that  the  Hamiltonian  has  2-spin
interactions.  Everything  we did so  far could then be  reproduced in
the  case  of $p$-spin  interactions,  and  we  would obtain  the  same
relations just derived, except the  overlaps would appear in the power
$p$ instead of  2.  Clearly the perturbation needed in  this case is a
$p$-spin  perturbation too.   More  in general,  we  could consider  a
Hamiltonian consisting of the  sum (over $p$) of $p$-spin Hamiltonians
for  any integer  $p$.  Then  we could  perturb each  of  the $p$-spin
Hamiltonians with its proper  small $p$-spin perturbation, and add all
these perturbations to  the system. Clearly we have  to make sure that
all the terms in this whole Hamiltonian are weighted with sufficiently
small weights so to  have the necessary convergence.  More explicitly,
the   perturbed   Hamiltonian   is   
$$
H_{N}(\sigma,\alpha;\mathcal{J})=
-\sum_{p}\bigg[a_{p}\sum_{\nu=1}^{P^{(p)}_{\alpha
    N}}J_{\nu}\sigma_{i^{1}_{\nu}}\cdots\sigma_{i^{p}_{\nu}}+b_{p}
\lambda_{p}\sum_{\nu=1}^{P^{\prime
    (p)}_{\alpha^{\prime}}}J^{\prime}_{\nu}\sigma_{j^{1}_{\nu}}
\cdots\sigma_{j^{p}_{\nu}}\bigg]\ ,          
$$
where
$\sum_{p}|a_{p}|^{2}=\sum_{p}|b_{p}|^{2}=1$, the  notation for all the
quenched  variables is the  usual one,  and $\{\lambda_{p}\}$  are the
independent perturbing real parameters.

It is not surprising then that we can state
\begin{corollary}\label{corollary}
With the possible exception of a zero measure set in the space
of all perturbing parameters, we have
$$
\sum_{a=0}^{\min\{2r,2s\}}(-)^{a+1}\frac{(2r+2s-a-1)!}{a!(2r-a)!(2s-a)!}
\langle q^m_{2r}q^n_{2s}\rangle^{\prime}_a=0 
\ \ \ \forall\ r,s,m,n \in \mathbb{N}\ .
$$
\end{corollary}
Again, this corollary can be seen as a consequence of 
a self-averaging property, namely
\begin{equation*}
\label{}
\mathbb{E}\Omega(\sigma_{i^1_1}\cdots\sigma_{i^m_1}
\sigma_{j^1_1}\cdots\sigma_{j^n_1})-
\mathbb{E}[\Omega(\sigma_{i^1_1}\cdots\sigma_{i^m_1})
\Omega(\sigma_{j^1_1}\cdots\sigma_{j^n_1})]=0\ .
\end{equation*}

Therefore we can replace each overlap by any smooth function
of the relative replicas in the statement of the corollaries.


\section{Self-averaging of the quenched-Gibbs measure}
\label{energysection}

Roughly speaking, if a convex random function does not
fluctuate much, then its derivative does not fluctuate much
either, with the exception of bad cases. This is well
explained in Proposition 4.3 of \cite{t5} and
Lemma 8.10 of \cite{bov2}. 
We are not interested
in general theorems, in our case the convex function
we are interested in is the free energy density, and 
we only need to know that it is self-averaging
(in the sense that the random free energy density
does not fluctuate around its quenched expectation,
in the thermodynamic limit). In the case
of finite connectivity random spin systems, a detailed proof
of this can be found in \cite{gt1}. The derivative
of the free energy density (times $-\beta)$ with respect to $-\beta$
is the expectation of the internal energy density $u_{N}=H_{N}/N$.
Like in \cite{guerra2} and in 
section 2 of \cite{gg}, 
we have therefore this further self-averaging
$$
\lim_{N\to\infty}[\langle u_{N}^{2}\rangle
-\langle u_{N}\rangle^{2}]=0
$$
which implies (due to Schwartz inequality)
\begin{equation}
\label{auto}
\lim_{N\to\infty}\langle u^{(1)}_{N}\phi_{s}\rangle
=\lim_{N\to\infty}\langle u_{N}\rangle\langle\phi_{s}\rangle
\end{equation}
for any bounded function  $\phi_{s}$ of $s$ replicas, and
$u^{(1)}_{N}$ is the internal energy density in the configuration space
of the replica 1. More precisely,
let us call the spin-configuration space $\{-1,1\}^{N}= \Sigma$,
and consider a bounded function $\phi_{s}$ of $s$ replicas, 
i.e. $\phi_{s}:\Sigma^{s}\to \mathbb{R}$.
The spin-configuration space $\Sigma$ is equipped with the 
Gibbs measure $\omega$,
and the product space $\Sigma^{s}$ (``the space of the replicas'')
is equipped with the product measure (``replica measure'')
$\omega^{\otimes s}=\Omega$. The quenched variables 
are the same in each factor of the product space, and this means
that the measure $\langle\cdot\rangle=\mathbb{E}\Omega(\cdot)
=\mathbb{E}\omega^{\otimes s}(\cdot)$
on the product space $\Sigma^{s}$ is not a product measure.
We will use for simplicity $\Omega$ for any value of $s$.
So $f^{(1)}_{N}$ is the free energy in the space which
is the first factor in the product space $\Sigma^{s}$.
Notice that $\Sigma$ has the cardinality of the continuum 
in the thermodynamic limit $N\to\infty$.
Apices will denumerate replicas for the spins and the
Hamiltonian, while they are just regular exponents in the case of 
overlaps, where the replicas are counted or listed in the sub-index. 

At this point 
we want to perturb the Hamiltonian and consider
the derivative with respect to the perturbing parameter,
as we did in the previous section:
$$
-\beta H_{N}(\sigma)\ \longrightarrow\ 
-\beta H_{N}(\sigma)+\beta^{\prime}\sum_{\nu=1}^{P^{\prime}\alpha}
J^{\prime}_{\nu}\sigma_{i^{\prime}_{\nu}}\sigma_{j^{\prime}_{\nu}} \ ,
$$
in order to obtain an expansion in powers $\beta^{\prime}$
with coefficients which do not depend on  $\beta^{\prime}$
in the thermodynamic limit.

We are going to prove, first of all, the following
\begin{theorem}\label{quattro} 
For a given bounded function $\phi_{s}$ of $s$ replicas,
the following relation constrains the distribution
of the 4-overlap
  \begin{multline*}
    \frac{s(s+1)(s+2)}{3!}
    \langle q^{2}_{1,s+1,s+2,s+3}\phi_{s}\rangle
    -\frac{s(s+1)}{2!}
    \sum_{a}^{2,s}\langle q^{2}_{1,a,s+1,s+2}\phi_{s}\rangle\\
    +s\sum_{a<b}^{2,s}\langle q^{2}_{1,a,b,s+1}\phi_{s}\rangle
    -\sum_{a<b<c}^{2,s}
    \langle q^{2}_{1,a,b,c}\phi_{s}\rangle
    =\langle q^{2}_{1234}\rangle\langle \phi_{s} \rangle \ .
  \end{multline*}
\end{theorem}
The proof is straightforward but long, and it will 
be splitted into several steps.

Let us consider the right hand side of (\ref{auto}). 
Put $t=\tanh(\beta^{\prime})$, $q_0=1$,
and let us just indicate the number
of replicas in the overlaps, rather than denumerating them all.
Recall also that $p_{N}=-\beta f_{N}$,
which here ``contains'' the perturbed Hamiltonian.
Let us prove the next
\begin{lemma}\label{energy} The derivative
of the (perturbed) pressure $p_N(\beta,\beta^{\prime})$
with respect to the perturbing parameter $\beta^{\prime}$
has the following form as a series in powers of
$t=\tanh(\beta^{\prime})$
  \begin{equation*}
    \partial_{\beta^{\prime}}p_{N}(\beta,\beta^{\prime})
    =-\alpha\sum_{n=0}^{\infty}
    t^{2n+1}(\langle q^{2}_{2n}\rangle-\langle q^{2}_{2n+2}\rangle)\ .
  \end{equation*}
\end{lemma}
{\bf Proof.} We have
\begin{eqnarray*}
  \partial_{\beta^{\prime}}p_{N}(\beta,\beta^{\prime})
  &=&-\sum_{m=1}^\infty
  \pi_{\alpha}(m)\sum_{\nu=1}^m\langle 
  J^{\prime}_\nu\sigma_{i^{\prime}_\nu}
  \sigma_{j^{\prime}_{\nu}}\rangle_m\\
  &=&-\sum_{m=1}^\infty
  m\pi_{\alpha}(m)\langle J^{\prime}_m\sigma_{i^{\prime}_m}
  \sigma_{j^{\prime}_{m}}\rangle_m\\
  &=& -\alpha\sum_{m=1}^\infty
  \pi_{\alpha}(m-1)\langle J^{\prime}_m\sigma_{i^{\prime}_m}
  \sigma_{j^{\prime}_{m}}\rangle_m
\end{eqnarray*}
where the sub $m$ indicates that the variable $P^{\prime}_{\alpha}$
has been fixed to $m$.
It is easy to see that
\begin{equation}\label{acca1}
  \langle J^{\prime}_m\sigma_{i^{\prime}_m}
  \sigma_{j^{\prime}_{m}}\rangle_m=
  \mathbb{E}\frac{\omega(J^{\prime}_m\sigma_{i^{\prime}_m}
  \sigma_{j^{\prime}_{m}}
    \exp(\beta J^{\prime}_m\sigma_{i^{\prime}_m}
    \sigma_{j^{\prime}_{m}}))_{m-1}}{\omega(
    \exp(\beta J^{\prime}_m\sigma_{i^{\prime}_m}
    \sigma_{j^{\prime}_{m}}))_{m-1}}\ .
\end{equation}
Hence
\begin{equation}\label{acca2}
  \partial_{\beta^{\prime}}p_{N}(\beta,\beta^{\prime})
  =-\alpha\mathbb{E}J^{\prime}
  \frac{t+w}{1+tw}\ ,\ w\equiv
  \omega(\sigma_{i^{\prime}_m}\sigma_{j^{\prime}_{m}})\ ,
\end{equation}
according to the usual notations. Now a simple expansion
(that we will explicitly write in the next lemma)
of $(1+tw)^{-1}$ in powers of $t$ yields
\begin{equation}\label{phioutside}
 \partial_{\beta^{\prime}}p_{N}(\beta,\beta^{\prime})
  =-\alpha\sum_{n=0}^{\infty}
  t^{2n+1}(\langle q^{2}_{2n}\rangle-\langle q^{2}_{2n+2}\rangle)\ .
\end{equation}
So the lemma is proven and we have an expression for
the right hand side of (\ref{auto}), if we just
multiply the average of the multi-overlaps by
the average of $\phi_s$.

Let us now consider the left hand side of (\ref{auto}),
recalling that $\phi_{s}$ is a function of $s$ replicas,
that indices in the spins indicate which factor of the
product space $\Sigma^{s}$ (which replica) the spin belongs to, 
and that the energy density is assumed
to be taken in the first replica.
We will henceforth omit the prime symbol in all
the quenched variables, but still assume that they
are independent of any other quenched variable
implicitly contained in the averages.
\begin{lemma}\label{espansione}
Recalling that $w\equiv\omega(\sigma_{i_m}
\sigma_{j_m})$, we have
  \begin{multline*}
     \langle u^{(1)}_{N} \phi_{s} \rangle=
    -\alpha t\mathbb{E}\{\Omega[\phi_{s}(1+Jt^{-1}\sigma^1_{i_1}
    \sigma^1_{j_1})\times \\
    (1+J\sum_a^{2,s}\sigma^a_{i_1}\sigma^a_{j_1} t+
    \sum_{a<b}^{2,s}\sigma_{i_1}^{a}\sigma_{i_1}^{b}
    \sigma_{j_1}^{a}\sigma_{j_1}^{b}t^2+
    \sum_{a<b<c}^{2,s}\sigma_{i_1}^{a}
    \sigma_{i_1}^{b}\sigma_{i_1}^{c}\sigma_{j_1}^{a}
    \sigma_{j_1}^{b}\sigma_{j_1}^{c}t^3
    +\cdots)]\times \\
    (1-Jstw+\frac{s(s+1)}{2!}t^{2}w^{2}-
    J\frac{s(s+1)(s+2)}{3!}t^3w^3\\
    +\frac{s(s+1)(s+2)(s+3)}{4!}t^4w^4- \cdots)\}\ .
  \end{multline*}
\end{lemma} 
{\bf Proof}. From the proof of the previous lemma, 
in particular equations (\ref{acca1})-(\ref{acca2}), and
by definition of replica measure, we immediately get
\begin{equation}\label{phiinside}
  \langle u^{(1)}\phi_{s} \rangle=
  -\alpha\mathbb{E}
  \frac{\Omega[ J\sigma^{1}_{i_1}\sigma^{1}_{j_1}
    \exp(\beta J(\sigma^1_{i_1}\sigma^{1}_{j_1}
    +\cdots+\sigma^s_{i_1}\sigma^{s}_{j_1}))\phi_{s}]}
  {\Omega^s(\exp (\beta J\sigma_{i_1}\sigma_{j_1}))}\ ,
\end{equation}
that we rewrite as
\begin{equation*}
  \langle u^{(1)}\phi_{s} \rangle=
  -\alpha\mathbb{E}t
  \frac{\Omega[(1+Jt^{-1}\sigma^1_{i_1}\sigma^1_{j_1})\prod_{a=2}^s
    (1+Jt\sigma^a_{i_1}\sigma^a_{j_1})\phi_{s}]}
  {(1+Jtw)^s}\ .
\end{equation*}
Let us write explicitly the power expansion of the denominator,
that we omitted in the previous lemma
\begin{multline*}
  \frac{1}{(1+Jtw)^s}=1-Jstw+
  \frac{s(s+1)}{2!}t^{2}w^{2}-\\
  J\frac{s(s+1)(s+2)}{3!}t^3w^3
  +\frac{s(s+1)(s+2)(s+3)}{4!}t^4w^4\cdots\ .
\end{multline*}
It is also clear that
\begin{multline*}
  \prod_{a=2}^s(1+Jt\sigma^a_{i_1}\sigma^{a}_{j_1})
  =1+J\sum_a^{2,s}\sigma^a_{i_1}\sigma^{a}_{j_1} t+
  \sum_{a<b}^{2,s}\sigma_{i_1}^{a}\sigma_{i_1}^{b}
  \sigma^{a}_{j_1}\sigma^{b}_{j_1}t^2\\
  +\sum_{a<b<c}^{2,s}\sigma_{i_1}^{a}
  \sigma_{i_1}^{b}\sigma_{i_1}^{c}\sigma^{a}_{j_1}
  \sigma^{b}_{j_1}\sigma^{c}_{j_1}t^3
  +\cdots\ .
\end{multline*}
Gathering all the ingredients completes the proof of the lemma.

We are now able to compare the two sides of (\ref{auto}),
and see what the self-averaging of the internal energy density
in the thermodynamic limit brings.

Equating the expressions computed in the 
last two lemmas gives
\begin{multline}\label{sviluppo}
  \sum_{n=0}^{\infty}
  t^{2n}(\langle q^{2}_{2n}\rangle
  -\langle q^{2}_{2n+2}\rangle)\langle \phi_{s} \rangle
  =\mathbb{E}\{\Omega[\phi_{s}(1+Jt^{-1}\sigma^1_{i_1}
  \sigma^{1}_{j_1})\\
  (1+J\sum_{a}^{2,s}\sigma^a_{i_1}\sigma^{a}_{j_1} t+
  \sum_{a<b}^{2,s}\sigma_{i_1}^{a}\sigma_{i_1}^{b}
  \sigma_{j_1}^{a}\sigma_{j_1}^{b}t^2+
  \sum_{a<b<c}^{2,s}\sigma_{i_1}^{a}
  \sigma_{i_1}^{b}\sigma_{i_1}^{c}\sigma_{i_1}^{a}\sigma_{i_1}^{b}
  \sigma^{c}_{j_{1}}t^3+\\
  \cdots+J^{s-1}t^{s-1}\sigma_{i_1}^2\cdots\sigma_{i_1}^s
  \sigma_{j_1}^2\cdots\sigma_{j_1}^s)]\\
  (1-Jstw+\frac{s(s+1)}{2!}t^2w^2-
  J\frac{s(s+1)(s+2)}{3!}t^3w^3\\
  +\frac{s(s+1)(s+2)(s+3)}{4!}t^4w^4-\cdots)\}\ .
\end{multline}
The equality holds for any smooth function $\phi_{s}$
(typical interesting information is obtained 
for $\phi_{s}\equiv1$ or $\phi_{s=2n}=q^{2}_{2n}$), so that we get
equalities between expressions
involving averages of (squared) overlaps. 

Let us see in detail what information we can get from the lowest orders.

Denote by $\mathbb{E}(\cdot|\mathcal{A}_s)$ the conditional
expectation with respect to the sigma-algebra $\mathcal{A}_s$
generated by the overlaps of $s$ replicas. Let us show that the usual
\cite{gg} Ghirlanda-Guerra identities for the overlap hold in our
quite general case too (as well known):
\begin{proposition} The Ghirlanda-Guerra relation holds 
  \begin{equation}\label{ghigu}
    \mathbb{E}(q^{2}_{a,s+1}|\mathcal{A}_s)=
    \frac1s\langle q^{2}_{12}\rangle
    +\frac1s\sum_{b\neq a} q^{2}_{a b}\ .
  \end{equation}
\end{proposition}
{\bf Proof}. In the expansion (\ref{sviluppo}), where only the terms of 
even order survive due to the symmetry of the variables $J$,
at the lowest order in $t$ one gets 
\begin{eqnarray*}
  \langle\phi_{s}\rangle-
  \langle q^{2}_{12}\rangle\langle\phi_{s}\rangle &=&
  \langle\phi_{s}\rangle-s\mathbb{E}
  [\omega(\sigma^1_{i_1}\sigma^1_{j_1})w\phi_{s}]
  +\sum_{a}^{2,s}\mathbb{E}[\Omega(
  \sigma^1_{i_1}\sigma^a_{i_1}\sigma^1_{j_1}\sigma^a_{j_1})\phi_{s}]\\
  {} &=& \langle\phi_{s}\rangle 
  -s\langle q^{2}_{1,s+1}\phi_{s}\rangle
  +\sum_{a}^{2,s}\langle q^{2}_{1a}\phi_{s}\rangle\ ,
\end{eqnarray*}
which is precisely what is stated in (\ref{ghigu}),
(see \cite{talabook}), immediately
completing the proof of the proposition.

So the usual Ghirlanda-Guerra
identities for 2-overlaps are recovered (and proven to hold in 
dilute spin glasses too, for instance).

At the next order we get instead
\begin{multline}\label{ordinedue}
  \langle q^{2}_{12}\rangle\langle\phi_{s}\rangle- \langle
  q^{2}_{1234}\rangle\langle\phi_{s}\rangle=
  \sum_{a<b}^{2,s}\langle q^{2}_{a b}\phi_{s}\rangle+
  \frac{s(s+1)}{2!}\langle q^{2}_{s+1,s+2}\phi_{s}\rangle\\
  -s\sum_a^{2,s}\langle
  q^{2}_{a,s+1}\phi_{s}\rangle-\frac{s(s+1)(s+2)}{3!}
  \langle q^{2}_{1,s+1,s+2,s+3}\phi_{s}\rangle\\
  +\frac{s(s+1)}{2!}\sum_a^{2,s}\langle
  q^{2}_{1,a,s+1,s+2}\phi_{s}\rangle- s\sum_{a<b}^{2,s}\langle
  q^{2}_{1,a,b,s+1}\phi_{s}\rangle\ +\sum_{a<b<c}^{2,s} \langle
  q^{2}_{1,a,b,c}\phi_{s}\rangle\ .
\end{multline}
Now consider the four 2-overlaps terms. 
A simple generalization of the usual Ghirlanda-Guerra relations
\cite{gg}
to the case when two replicas are added to a previously assigned
set of other replicas, tells us that these terms cancel out.
Let us check that explicitly.
\begin{corollary} Relation (\ref{ghigu}) implies
  \begin{equation}\label{ghigu2}
    \mathbb{E}(q^{2}_{s+1,s+2}|\mathcal{A}_{s}) =
    \frac{2}{s+1}\langle q^{2}_{12}\rangle +
    \frac{2}{s(s+1)}\sum_{a<b}^{1,s}q^{2}_{ab}\ .
  \end{equation}
\end{corollary}
{\bf Proof}.
Let us re-write (\ref{ghigu}) in the case of $s+1$ given
replicas
$$
\mathbb{E}(q^{2}_{s+1,s+2}|\mathcal{A}_{s+1})
=\frac{1}{s+1}\langle q^{2}_{12}\rangle
+\frac{1}{s+1}\sum_{b}^{1,s} q^{2}_{b,s+1}\ .
$$
Now use 
\begin{equation}\label{sigma}
  \mathbb{E}(\mathbb{E}(\cdot|\mathcal{A}_{s+1})|\mathcal{A}_s)
  =\mathbb{E}(\cdot|\mathcal{A}_s)
\end{equation}
to get
\begin{eqnarray*}
  \mathbb{E}(q^{2}_{s+1,s+2}|\mathcal{A}_{s}) & = & 
  \frac{1}{s+1}\langle q^{2}_{12}\rangle
  +\frac{1}{s+1}\sum_{b}^{1,s} \mathbb{E}(q^{2}_{b,s+1}|\mathcal{A}_s)\\
  {} & = & \frac{1}{s+1}\langle q^{2}_{12}\rangle +
  \frac{1}{s+1}\left(\langle q^{2}_{12}\rangle + 
    \frac1s \sum_{b}^{1,s}\sum_{c\neq b}^{1,s}q^{2}_{bc}\right)\ .
\end{eqnarray*}
That is
\begin{equation*}
  \mathbb{E}(q^{2}_{s+1,s+2}|\mathcal{A}_{s}) =
  \frac{2}{s+1}\langle q^{2}_{12}\rangle +
  \frac{2}{s(s+1)}\sum_{a<b}^{1,s}q^{2}_{ab}\ ,
\end{equation*}
which is what we wanted to prove.

Now with (\ref{ghigu}) and (\ref{ghigu2}) in our hands, let us 
take the three 2-overlap terms in the right hand side of (\ref{ordinedue})
\begin{eqnarray*}
  \frac{s(s+1)}{2}\langle q^{2}_{s+1,s+2}\phi_{s}\rangle & = &
  s\langle q^{2}_{12}\rangle\langle\phi_{s}\rangle 
  +\sum_{a<b}^{1,s}\langle q^{2}_{ab}\phi_{s}\rangle \\
  -s\sum_{a}^{2,s}\langle q^{2}_{a,s+1}\phi_{s}\rangle & = & 
  -s\sum_{a}^{1,s}\langle q^{2}_{a,s+1}\phi_{s}\rangle 
  + s\langle q^{2}_{1,s+1}\phi_{s}\rangle \\
  {} & = & -s\langle q^{2}_{12}\rangle\langle\phi_{s}\rangle
  -\sum_{a}^{1,s}\sum_{b\neq a}^{1,s}\langle q^{2}_{ab}\phi_{s}\rangle 
  +\langle q^{2}_{12}\rangle\langle\phi_{s}\rangle 
  +\sum_{a}^{2,s}\langle q^{2}_{1a}\phi_{s}\rangle\\
  \sum_{a<b}^{2,s}\langle q^{2}_{ab}\phi_{s}\rangle & = &
  \sum_{a<b}^{1,s}\langle q^{2}_{ab}\phi_{s}\rangle-
  \sum_{a}^{2,s}\langle q^{2}_{1a}\phi_{s}\rangle\ .
\end{eqnarray*}
The sum of these three terms cleary reduces to 
$\langle q^{2}_{12}\rangle\langle\phi_{s}\rangle$, which is precisely
what we find in the left hand side of (\ref{ordinedue}).
The 2-overlap terms thus cancel out from (\ref{ordinedue}). 
We are hence left with a new relation for 4-overlaps:
\begin{multline*}
  \frac{s(s+1)(s+2)}{3!}
  \langle q^{2}_{1,s+1,s+2,s+3}\phi_{s}\rangle
  -\frac{s(s+1)}{2!}
  \sum_{a}^{2,s}\langle q^{2}_{1,a,s+1,s+2}\phi_{s}\rangle\\
  +s\sum_{a<b}^{2,s}\langle q^{2}_{1,a,b,s+1}\phi_{s}\rangle
  =\langle q^{2}_{1234}\rangle\langle \phi_{s} \rangle
  +\sum_{a<b<c}^{2,s}
  \langle q^{2}_{1,a,b,c}\phi_{s}\rangle\ ,
\end{multline*}
and the proof of Theorem \ref{quattro} is now complete.


We report for sake of completeness the general
expression of the generic order in the power series
expansion (\ref{sviluppo}). From the explicit calculation
in Lemma \ref{espansione} we get
\begin{multline*}
\langle q^{2}_{2n}\rangle\langle\phi_{s}\rangle
-\langle q^{2}_{2n+2}\rangle\langle\phi_{s}\rangle=\\
\sum_{m=2n-s+1}^{2n}\sum_{l=0}^{s-1}\sum_{a_1<\cdots<a_l}^{2,s}
(-)^m\binom{s+m+1}{m}
\mathbb{E}[w^m\Omega(\phi_{s}
\sigma^{a_1}_{i_1}\cdots\sigma^{a_l}_{i_1}
\sigma^{a_1}_{j_1}\cdots\sigma^{a_l}_{j_1})]\delta_{2n,m+l}\\
+\sum_{m=2n-s+2}^{2n+1}\sum_{l=0}^{s-1}\sum_{a_1<\cdots<a_l}^{2,s}
(-)^m\binom{s+m+1}{m}
\mathbb{E}[w^m\Omega(\phi_{s}\sigma^1_{i_1}\sigma^1_{j_1}
\sigma^{a_1}_{i_1}\cdots\sigma^{a_l}_{i_1}
\sigma^{a_1}_{j_1}\cdots\sigma^{a_l}_{j_1})]\delta_{2n,m+l-1}
\end{multline*}
which becomes
\begin{multline}\label{recursive}
\langle q^{2}_{2n}\rangle\langle\phi_{s}\rangle
-\langle q^{2}_{2n+2}\rangle\langle\phi_{s}\rangle=
\sum_{l=0}^{2n \land s-1}\sum_{a_1<\cdots<a_l}^{2,s}
(-)^{2n-l}\binom{2n+s-l+1}{2n-l}\times \\
[\langle \phi_{s}
q^{2}_{a_1\cdots a_{l}}q^{2}_{s+1\cdots s+2n-l}\rangle
-\frac{2n-l+s+2}{2n-l+1}
\langle\phi_{s} q^{2}_{1a_1\cdots a_{l}}q^{2}_{s+1\cdots s+2n-l+1}
\rangle]\  .
\end{multline}
In both the expressions above the term for $l=0$ is understood to be one.

The right hand side of (\ref{recursive}), due to the presence of 
$1+Jt^{-1}\sigma$ in the 
right hand side of (\ref{sviluppo}) - along with the 
symmetry of $J$, makes the expansion
somewhat recursive. This means that at each order we find some terms
already found in the previous order. More precisely, we claim
without proving that at each $2n$-th order 
of the expansion, all the terms involving $2m$-overlaps with $2m\leq 2n$
cancel out thanks to a repeated use of (\ref{sigma}) with 
the relations coming from the lower orders. Hence
from the $2n$-th order we get new relations involving $2n+2$-overlaps
only. This is what we explicitly verified only for 4-overlaps
in the previous pages.
More explicitly, if we re-write the difference in the right hand side  
of (\ref{recursive}) as
$$
\langle q^{2}_{2n}\rangle\langle\phi_{s}\rangle
-\langle q^{2}_{2n+2}\rangle\langle\phi_{s}\rangle=
c_{2n}-d_{2n+2}\ ,
$$
we have
$$
\langle q^{2}_{2n}\rangle\langle\phi_{s}\rangle=c_{2n}\ ,\ 
\langle q^{2}_{2n+2}\rangle\langle\phi_{s}\rangle=d_{2n+2}\ ,\
c_{2n}=d_{2n}\ .
$$
So that the final formula becomes
\begin{multline*}
\langle q^{2}_{2n}\rangle\langle\phi_{s}\rangle=\\
\sum_{l=0}^{2n \land s-1}\sum_{a_1<\cdots<a_l}^{2,s}
(-)^{2n-l}\binom{2n+s-l+1}{2n-l}
\langle q^{2}_{a_1\cdots a_{l}}q^{2}_{s+1\cdots s+2n-l}
\phi_{s}\rangle \ .
\end{multline*}


\subsection{generalization to smooth functions of 
multi-overlaps}

Just like for the family of identities discussed in the previous section,
we started our analysis with the most natural quantity: the energy
of our model with 2-spin interactions.
And so we got again some relations for the squared multi-overlaps.
But we already know how to generalize these formulas to 
smooth functions of the overlaps. We can consider $p$-spin interactions,
and the procedure would provide us with the same relations for the 
$p$-th power of the overlaps. Then, as already explained, we can
take a convergent sum over all integer $p$ of $p$-spin Hamiltonians,
and consider the self-averaging of the desired one among them.
The perturbed Hamiltonian is again
$$
H_{N}(\sigma , \alpha ; \mathcal{J})=-\sum_{p}\bigg[a_{p}
\sum_{\nu=1}^{P^{(p)}_{\alpha N}}J_{\nu}\sigma_{i^{1}_{\nu}}
\cdots\sigma_{i^{p}_{\nu}}
+b_{p}\lambda_{p}\sum_{\nu=1}^{P^{\prime (p)}_{\alpha^{\prime}}}
J^{\prime}_{\nu}
\sigma_{j^{1}_{\nu}}
\cdots\sigma_{j^{p}_{\nu}}\bigg]\ ,
$$
where $\sum_{p}|a_{p}|^{2}=\sum_{p}|b_{p}|^{2}=1$, 
the notation for all the quenched variables is the usual one,
and $\{\lambda_{p}\}$ are the independent perturbing
real parameters. As a side remark, we just point out that
(like in \cite{gg}), in the case of this secon family
of identities it is not necessary to consider a 
Hamiltonian consisting of the sum of all possible
$p$-spin Hamiltonians: only the perturbation must be
so.


\section*{Concluding remarks}

Notice that while we derived our identities having as reference 
diluted spin glasses, all that matters in the derivation are the
properties of the perturbing Hamiltonian, and they are therefore
generically valid.

The Ghirlanda-Guerra identities for the overlap have been useful to
prove non trivial properties of mean-field spin glasses. For instance
Talagrand could prove that for all models where the identities are
valid, the support of the overlap probability function has positive
support. This positivity property is important as it enters in the the
Guerra free-energy bounds in spin system without spin reversal
symmetry. The corresponding bounds for diluted systems involve all
possible multioverlap. It has been proved \cite{flt} that the cavity
method provides free-enegy lower bounds for the random K-SAT problem
for even K. Due to the difficulty of proving the positivity of the
multioverlap, the bound does not apply to the odd K case. Proving the
positivity would therefore allow to extend the bound to this case and
in particular to the symbolic case K=3. Unfortunately the derivation
of Talagrand for the overlap does not extend immediately to the
multi-overlap case. We believe however that the self-averaging
identity will be useful in the mathematical analysis of diluted spin
models.

\section*{Acknowledgments}

LDS thanks Fabio Lucio Toninelli and Anton Bovier for useful discussions.


\end{document}